\documentclass[11pt,preprintnumbers,aps,amssymb,nofootinbib,amsmath,superscriptaddress,notitlepage,prd]
{revtex4-1}
\usepackage{epsfig,epsf}
\usepackage{bm} 
\usepackage{color} 
\definecolor{darkgreen}{RGB}{0,128,0}
\definecolor{darkred}{RGB}{139,0,0}
\usepackage{slashed} 
\usepackage{hyperref}
\usepackage{ulem}
\newcommand{\beq}{\begin{equation}}
\newcommand{\beql}[1]{\begin{equation}\label{#1}}
\newcommand{\eeq}{\end{equation}}
\def\bal#1\gal{\begin{align}#1\end{align}}
\newcommand{\ball}[1]{\bal\label{#1}}
\newcommand{\eq}[1]{(\ref{#1})}

\renewcommand{\b}[1]{{\bm #1}}

\newcommand{\e}{\varepsilon}

\newcommand{\sgn}{\operatorname{sgn}}

\begin{document}

\title{Transition radiation as a probe of chiral anomaly}

\author{Xu-Guang Huang}
\affiliation{Physics Department and Center for Particle Physics and Field Theory, Fudan University, \\ Shanghai 200433, China}
\affiliation{Key Laboratory of Nuclear Physics and Ion-beam Application (MOE), Fudan University, \\ Shanghai 200433, China}

\author{Kirill Tuchin}

\affiliation{Department of Physics and Astronomy, Iowa State University, Ames, IA 50011, USA}

\date{\today}


\begin{abstract}

A fast charged particle crossing the boundary between the chiral matter and vacuum radiates the transition radiation. Its most remarkable features --- the resonant behavior at a certain emission angle and  the circular polarization of the spectrum --- depend on the parameters of the chiral anomaly in a particular material/matter. The chiral transition radiation can be used to investigate the chiral anomaly in such diverse media  as the quark-gluon plasma, the Weyl semimetals, and the axionic dark matter.

\end{abstract}

\maketitle

\section{Introduction}

The chiral matter --- the matter containing chiral fermions --- possesses a number of unique properties originating from the quantum phenomenon of the chiral anomaly. Those chiral materials that exist at room temperatures, such as the Weyl semimetals, can be studied with high precision. On the other hand, there are forms of chiral matter that exist only under extreme conditions, such as the quark-gluon plasma; their study requires novel approaches. In this letter we argue that an informative insight into properties of the chiral matter can be gained using the chiral analogue of the transition radiation.

The transition radiation is emitted when a fast charged particle, i.e.\ a particle moving with energy much greater than the medium ionization energy, crosses the boundary between the two media having different dielectric constants. This is a classical effect predicted by Ginzburg and Frank in 1945 \cite{Ginzburg:1945zz} (reviewed in \cite{Ginzburg-Tsytovich}) that has a number of practical applications. The quantum corrections were calculated in \cite{Garibyan,Baier:1998ej,Schildknecht:2005sc}. The transition radiation originates  from the difference of the photon wave function on the two sides of the boundary. At high energies this is manifested in variation of the plasma frequency, across the boundary.

 In a chiral matter photon dispersion relation is modified due to the chiral anomaly \cite{Deser:1981wh}. As a result, when a fast charged particle crosses the boundary between the chiral matter and vacuum it emits the transition radiation, which we will refer to as the {\it chiral transition radiation}. Its spectrum was recently derived by one of us  in \cite{Tuchin:2018sqe} employing a method developed in \cite{Schildknecht:2005sc}. It possesses distinctive features as compared to other forms of radiation by fast particles in matter. Thus, the chiral transition radiation can be employed to investigate the chiral anomaly in various forms of matter and materials as we explain in the forthcoming sections.

\section{The spectrum}

The dispersion relation of a photon in a chiral medium can be most readily computed using the Maxwell-Chern-Simons theory \cite{Wilczek:1987mv,Carroll:1989vb,Sikivie:1984yz,Kalaydzhyan:2012ut}, which is an effective low energy approximation of the QED in chiral medium. The gauge part of this theory reads
\ball{a1}
\mathcal{L}= -\frac{1}{4}F_{\mu\nu}^2-\frac{c_A}{4}\theta\tilde F_{\mu\nu}F^{\mu\nu}\,,
\gal
where the pseudo-scalar field $\theta$ encapsulates the effect of the chiral anomaly  and $c_A$ is the anomaly coefficient. In practical applications one usually assumes that $\theta$ is either  (i)  spatially uniform and adiabatically time dependent $\dot \theta\neq 0$, or (ii) that it is time-independent and slightly anisotropic $\b\nabla\theta \neq 0$. The dispersion relation in each case takes form \cite{Tuchin:2014iua,Tuchin:2017vwb,Yamamoto:2015maz,Qiu:2016hzd}
\ball{a3}
\omega^2= \b k^2+ \mu^2(\b k, \lambda)\,,
\gal
where $\b k$ is the photon momentum and $\lambda=\pm 1$ is its circular polarization. The parameter $\mu$ is  a complex function of its arguments that is sensitive to the spatial or temporal variation of the $\theta$-field.  In the case (i) it reads 
\ball{a6}
\mu^2(\b k, \lambda)=-\lambda \sigma_\chi k\,,
\gal
where $\sigma_\chi = c_A\dot \theta$ is  the chiral conductivity\cite{Fukushima:2008xe,Kharzeev:2009pj}. In the case (ii) it takes form
\ball{a5}
\mu^2(\b k, \lambda)=\frac{1}{2}b^2-\lambda \sgn(\b k\cdot \b b)\sqrt{(\b k\cdot \b b)^2+\frac{1}{4}b^4} \,,
\gal
where  $\b b= c_A\b\nabla\theta$ \cite{Qiu:2016hzd}.  In Weyl semimetals $\b\nabla \theta $  is the separation in momentum space between the Weyl nodes of right-handed and left-handed fermions. We observe that   $\mu$ can be real or imaginary  depending on the photon polarization, whereas in a non-chiral matter $\mu$ is always real. This is the origin of the distinct transition radiation pattern from the chiral matter that we are proceeding to discuss in the next few paragraphs.

We start with the case (i) representing spatially uniform matter.  We assume that  the boundary is located at $z=0$ and the particle moves in the $z$ direction, i.e.\ perpendicular to the boundary. At the boundary $\mu$ is discontinuous. In the ultrarelativistic limit, when $\mu$ can be treated as a small parameter, the photon wave function in the radiation gauge reads 
\ball{a8}
\b A= \frac{1}{\sqrt{2\omega V}}\b  \epsilon_\lambda  \,e^{i \omega z+i\b k_\bot \cdot \b x_\bot-i\omega t }\exp\left\{ -i \frac{1}{2\omega} \int_0^z(k_\bot^2+\mu^2)dz'\right\}\,,
\gal
where  $\b  \epsilon_\lambda$ is the polarization vector  such that $\b  \epsilon_\lambda \cdot \b k=0$ and $V$ is the normalization volume. By the same token the fermion wave function is
\ball{a9}
\psi = \frac{1}{\sqrt{2\e V}}u(p)  e^{i\e z-i \e t}\exp\left\{ i\b p_\bot\cdot \b x_\bot-iz\frac{\b p_\bot^2+m^2}{2\e}\right\}\,,
\gal
where $\b p$ and $\e$ are the fermion momentum and energy.

The scattering matrix element for the photon emission process is
\bal
S=& -ie Q\int \bar \psi \gamma^\mu \psi A_\mu d^4x=i(2\pi)^3\delta(\omega+\e'-\e)\delta(\b p_\bot-\b k_\bot-\b p'_\bot)\frac{\mathcal{M}}{\sqrt{8\e\e' \omega V^3}}\,,
\label{a13}
\gal
where $Q$ is the fermion electric charge and the prime distinguishes the final fermion energy and momentum. The invariant amplitude reads 
\ball{a15}
\mathcal{M}&=  -eQ\bar u(p')\slashed{\epsilon}^* u(p)\, 2\e x(1-x)\left\{ \frac{-i}{q_\bot^2+\kappa_\lambda-i\gamma}-\frac{-i}{q_\bot^2+x^2m^2+i\gamma}
\right\}
\,,
\gal
where $x=\omega/\e$ is the fraction of the incident fermion energy carried away by the radiated photon,  $\b q_\bot = x\b p-\b k_\bot$, $\kappa_\lambda = x^2m^2+(1-x)\mu^2$ and $\gamma$ is the resonance width  that depends on the system geometry, electrical conductivity etc. The radiated photon spectrum can be computed as
\ball{a19}
\frac{dN}{d^2q_\bot dx}= \frac{1}{(2\pi)^3}\frac{1}{8x(1-x)\e^2}\frac{1}{2}\sum_{\lambda,\sigma,\sigma'}|\mathcal{M}|^2\,,
\gal
where $\sigma$,$\sigma'$ are the fermion and anti-fermion spins. Substitution of \eq{a19} into \eq{a15} yields
\ball{a21}
\frac{dN}{d^2q_\bot dx}= \frac{\alpha Q^2}{2\pi^2 x}\left\{ \left(\frac{x^2}{2}-x+1\right)q_\bot^2+\frac{x^4m^2}{2}\right\}
\sum_\lambda \left| \frac{1}{q_\bot^2+\kappa_\lambda-i\gamma}-\frac{1}{q_\bot^2+x^2m^2+i\gamma}\right|^2\,.
\gal
For positive $\kappa_\lambda$, the photon spectrum \eq{a21} coincides with the standard formula for the transition radiation with $\mu$ being the plasma frequency \cite{Schildknecht:2005sc}. However, the main contribution to the photon spectrum arises from the pole at $q_\bot^2= -\kappa_\lambda>0$, i.e.\ when $\kappa_\lambda$ is negative. Keeping only the term that is most singular at $\gamma\to 0$, we find the chiral transition radiation spectrum of photons \cite{Tuchin:2018sqe}
\ball{a24}
\frac{dN}{d^2q_\bot dx}=\frac{\alpha Q^2}{2\pi^2 x}\left\{ \left(\frac{x^2}{2}-x+1\right)q_\bot^2+\frac{x^4m^2}{2}\right\}
\frac{1}{(q_\bot^2+\kappa_\lambda)^2+\gamma^2}\,.
\gal
It is remarkable that the spectrum is  circularly polarized\footnote{In contrast, the ordinary transition radiation is linearly polarized \cite{Ginzburg:1945zz}.}. Indeed, $\kappa_\lambda$ is negative only if $\lambda \sigma_\chi>0$ and $x<[1+m^2/(\lambda\sigma_\chi\e)]^{-1}$. In other words, only one of the possible photon polarizations exhibits the resonant behavior, while the other one is suppressed.  Whether the photon spectrum is right- or left-hand polarized depends on the sign of $\sigma_\chi$.

Since $\mu^2\approx -\lambda\sigma_\chi \omega$,  the angular distribution of the photons peaks at the angle $\vartheta^2= q^2_\bot/ \omega^2 = -\kappa_\lambda/ x^2\e^2$ with respect to the  fermion momentum. If the fermion mass is negligible and bearing in mind that most photons are soft ($x\ll 1$) we can estimate $\vartheta^2\approx\lambda\sigma_\chi/ \omega$. 

\section{Applications}

{\bf 1.} As the first application, consider jet emission from the quark-gluon plasma (QGP) with a homogenous chiral conductivity. QGP is isotropic at the scales of interest here, hence the corresponding  case is  (i). Jets in heavy-ion collisions are produced by the highly energetic color particles. If a jet is originated by a quark (as opposed to a gluon) we  expect radiation  of circularly polarized photons in a cone with the opening angle  $\vartheta\sim \sqrt{|\sigma_\chi|/ \omega}$ with respect to the jet momentum. The chiral conductivity is an unknown parameter. If we estimate it as $\sigma_\chi \sim 10$~MeV, then $\omega =1$~GeV photons are emitted at the angle  $\vartheta\sim 0.1$, provided that the jet energy $\e$ is much larger than $\omega$.  Thus the observation of  circularly polarized photons at angle $\vartheta$ to the jet direction would be an indication of the chiral transition radiation.

{\bf 2.}  We have seen that the main feature of the transition radiation from chiral matter is the emergence of the resonance factor in \eq{a24}. It arises entirely due to the energy and momentum conservation in a $1\to 2$ process involving  a photon with complex $\mu$. Thus we expect to see  the same resonant factor as in  \eq{a24} arising in the case (ii) which deals with an anisotropic matter. The calculation of the pre-factor requires a more careful analysis that
 will be presented elsewhere. In the high energy limit Eq.~\eq{a5} reduces to $\mu^2\approx -\lambda \omega b\cos\beta$, where $\beta$ is the angle between $\b b$ and the photon momentum. The soft photon emission angle in the massless limit is $\vartheta^2 \approx \lambda b \cos\beta/\omega$.  Similarly to the previous case (i), the photon spectrum is circularly polarized. One can verify that now $\kappa_\lambda$ is negative only if $\lambda\cos\beta>0$ and $x< [1+m^2/(\lambda\e b\cos\beta)]^{-1}$.  Thus the polarization direction depends on whether $\b b$ points towards or away from  the boundary.
Furthermore, since $\mu^2$ is proportional to $\cos\beta$, the radiation is maximal when $\beta =0$ or $\pi$ and vanishes in the perpendicular direction. To estimate the characteristic radiation angle discussed above, consider a Weyl semimetal with  $b= (\alpha/\pi) 80$~eV \cite{Xu:2015cga,Lv:2015pya}. An electron with energy about GeV moving parallel to $\b b$ ($\beta=0$) would radiate, say, $\omega= 10$~MeV photons at $\vartheta=1.3\cdot 10^{-4}$. 
This can be tested by injecting a beam of energetic electrons normal to a Weyl semimetal film and measuring the polarization and angular distribution of the photons emitted in a cone with the opening angle $\vartheta$ around the beam direction.

{\bf 3.} The chiral transition radiation emitted by protons traveling through the dark matter lumps  \cite{Chang:2015odg} can be used to search for the axionic dark matter.  In this case $\dot\theta$ is proportional to the axion mass $m_a$ which is unknown but expected to be very small. The emission angle  of the chiral transition radiation with respect to the direction of a cosmic ray is of the order of $\sqrt{c_A\theta_0m_a/\omega}$  where $\theta_0$ is the average value of $\theta$. Taking $\theta_0\sim 10^{-19}$~\cite{Graham:2013gfa}, $m_a\sim 10^{-6}$~eV, and $\omega\sim 1$~TeV we obtain $\vartheta\sim  10^{-15}$.  Measurement of photon spectrum emitted by a cosmic ray at such angles might be possible   over the astronomical distances. 
\section{Summary}

In summary, we computed the transition radiation spectrum at the boundary between chiral matter and vacuum, given by \eq{a24} and argued that its unique features --- the resonant enhancement at a characteristic angle $\vartheta$ and  circular polarization --- can be used as the direct measurement of the chiral anomaly in chiral matter/materials.

\acknowledgments
We thank Dima Kharzeev for valuable discussions. XGH is supported by the Young 1000 Talents Program of China, NSFC through Grants No.\ 11535012 and No.\ 11675041. KT is supported in part by the U.S. Department of Energy under Grant No.\ DE-FG02-87ER40371.


\end{document}